# A Method for Parsing and Vectorization of Semi-structured Data used in Retrieval Augmented Generation

Hang Yang[1], Jing Guo[1], Jianchuan Qi[1], Jinliang Xie[1], Si Zhang[2], Siqi Yang[1], Nan Li[1,†], Ming Xu[1,*]

[1]Tsinghua University, [2]Minzu University of China

[†]li-nan@tsinghua.edu.cn, [*]xu-ming@mail.tsinghua.edu.cn,

## Abstract

This paper presents a novel method for parsing and vectorizing semi-structured data to enhance the functionality of Retrieval-Augmented Generation (RAG) within Large Language Models (LLMs). We developed a comprehensive pipeline for converting various data formats into .docx, enabling efficient parsing and structured data extraction. The core of our methodology involves the construction of a vector database using Pinecone, which integrates seamlessly with LLMs to provide accurate, context-specific responses, particularly in environmental management and wastewater treatment operations. Through rigorous testing with both English and Chinese texts in diverse document formats, our results demonstrate a marked improvement in the precision and reliability of LLMs outputs. The RAG-enhanced models displayed enhanced ability to generate contextually rich and technically accurate responses, underscoring the potential of vector knowledge bases in significantly boosting the performance of LLMs in specialized domains. This research not only illustrates the effectiveness of our method but also highlights its potential to revolutionize data processing and analysis in environmental sciences, setting a precedent for future advancements in AI-driven applications. Our code is available at https://github.com/linancn/TianGong-AI-Unstructure.git.

**Keywords**：Unstructured Data；Large Language Model；RAG；Partitioning；Chunking

## 1 Introduction

Large Language Models (LLMs) present substantial benefits in various specialized fields, particularly due to their proficiency in processing and deriving insights from extensive volumes of unstructured text. These models excel in converting intricate, unstructured data into organized formats, which is crucial for tasks such as predicting reaction conditions in scientific studies or isolating pertinent legal clauses from extensive documents. This capability is invaluable, especially for augmenting experimental databases and melding computational and experimental data, with notable applications in environmental science(Rillig et al., 2023). In the medical sector, LLMs have shown remarkable efficacy in named entity recognition (NER) tasks, facilitating the extraction and categorization of biomedical information from expansive data sets(Lee et al., 2020). This has significantly contributed to both research and clinical practice. Similarly, in the legal realm, LLMs have proven effective in analyzing complex legal documents, pinpointing crucial legal terms, and enhancing contract analysis(L. Yue et al., 2024). These applications underscore the transformative impact of LLMs in processing large and complex datasets into actionable insights, thus optimizing operations in specialized domains such as healthcare and law.

However, the integration of LLMs in specialized domains still faces challenges(Peng et al., 2023.). A notable issue is the generation of 'hallucinations' (L. Yang et al., 2024),which means the creation of factually incorrect, yet seemingly plausible information. This problem is compounded when addressing highly specialized or nuanced queries within professional contexts. This limitation predominantly originates from the generalized nature of the datasets used to train these models, which often lack the depth and specificity required for particular legal and medical scenarios(S. Pan et al., 2024). Consequently, this underscores the critical need for a strategic integration of LLMs with domain-specific expertise. Such a fusion, complemented by continuous evaluation and refinement, is essential to ensure the accuracy and relevance of the models' outputs, especially in fields where precision is paramount.

In the realm of ecological environmental management, the Retrieval-Augmented Generation (RAG) approach is highly relevant for LLMs applications. RAG integrates the capabilities of LLMs with external databases, enabling access to and incorporation of essential data during generation. This enhances the model's ability to provide accurate, context-specific information, crucial in environmental management's complex domain. However, implementing RAG faces significant challenges, notably in developing a vector-based knowledge base essential for accurate data retrieval. The complexity of creating this base from vast, unstructured environmental data is compounded by a lack of efficient structuring methods. Addressing these data processing challenges is imperative to fully utilize RAG's potential, thereby improving LLMs' effectiveness in ecological environmental governance.

In this study, we present an efficient method for processing documents in the `.docx` format and constructing a vector database, leveraging an unstructured open-source toolkit, the function calling capacity of OpenAI and the vector database platform of Pinecone. This paper details the method and their application in processing professional books for wastewater treatment plant operation and constructing a vector database for use with Retrieval-Augmented Generation (RAG), aiming to improve the expertise of large language models in the domain of wastewater treatment plant operation.

## 2 Background and Related work

Retrieval Augmented Generation (RAG) within large language models (LLMs) marks a significant stride in AI research, blending advanced knowledge retrieval with the generation capabilities of LLMs. This approach aims to boost the accuracy and relevance of the models' responses while preserving their contextual depth. Current research focuses on fine-tuning the retrieval process, ensuring that the information fetched aligns closely with user queries and enhances the quality of the model's output(Lewis et al., 2021.). A key challenge lies in integrating this retrieved information smoothly into the generation process, creating responses that are both coherent and contextually appropriate(Rohde et al., 2021).

A significant area of exploration is in improving the retrieval phase to filter out irrelevant information or 'noise', ensuring that the data used by the model is of high quality and relevance(Karpukhin et al., 2020). Researchers are also working on making LLMs more adaptable in using this retrieved data across various topics, enhancing the algorithms that control how the model accesses and uses this information(Kalyan et al., 2021).

Central to RAG's function in LLMs is the creation of vector databases from unstructured or semi-structured data like texts and web pages. These databases store information in a format that LLMs can easily access and use. Current research, including work on Transformer-based models, is pivotal in developing methods to efficiently transform vast amounts of data into these useful vector formats (Devlin et al., 2019).

However, a noticeable gap in this area is the lack of simple, efficient methods for creating these vector databases. Existing techniques, while effective, tend to be complex and resource-heavy, limiting their broader application. Addressing this challenge with more user-friendly vectorization methods is crucial. Such advancements would significantly widen the scope and effectiveness of LLMs, enabling them to process and generate more nuanced, context-rich language responses in a range of fields, thus enhancing the practical utility and reach of LLMs in various applications.

## 3 Core Functions

However, a noticeable gap in this area is the lack of simple, efficient methods for creating these vector databases. Existing techniques, while effective, tend to be complex and resource-heavy, limiting their broader application. Addressing this challenge with more user-friendly vectorization methods is crucial. Such advancements would significantly widen the scope and effectiveness of LLMs, enabling them to process and generate more nuanced, context-rich language responses in a range of fields, thus enhancing the practical utility and reach of LLMs in various applications.

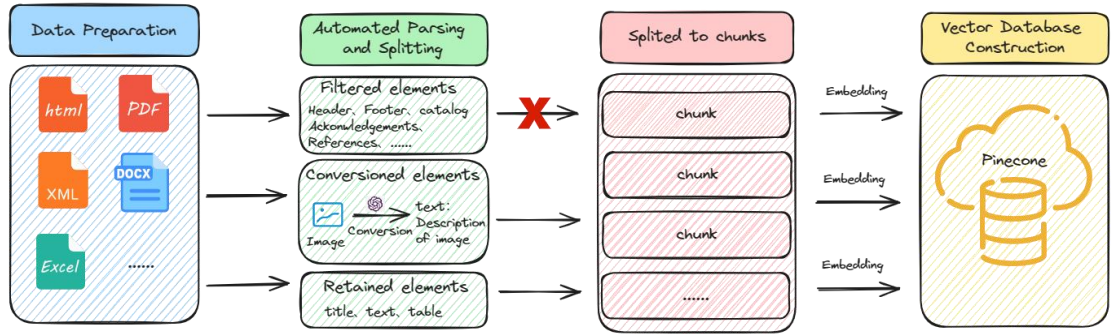

**Fig. 1 Parsing and Vectorization of Semi-structured Data process framework**

## 3.1 Data Preparation

In this phase, a diverse array of sources including books, reports, scholarly articles, and data tables is compiled.These data largely consists of semi-unstructured data, encompassing a variety of file formats such as `.html`, `pdf`, `xml`, `docx`, `xlsx` and etc. Considering the substantial volume of data to be processed, the `.docx` format stands out due to its uniform standardization, high-quality text, ease of editing, broad compatibility, and rich metadata content, making it highly advantageous for efficient bulk processing and structured data extraction.In this project, API functionalities are employed to integrate open-source tools for the purpose of converting diverse data formats into the .docx format.

For the assurance of effective post-processing, it is imperative that the content in the transformed `.docx` files, including headings, textual elements, and tables, be conformed to a standardized format. This standardization process involves harmonizing the font type, font size, inter-paragraph spacing, and line spacing across all headings, main text, and table contents.

## 3.2 Automated parsing and splitting

During the parsing process, the `.docx` files are divided into multiple elements including titles, texts, images, tables, headers and footers with the partitioning function, utilizing detectron2, a deep learning-based object detection system (Unstructured, 2023). This partition function uses a combination of the styling information in the document and the structure of the text to determine the type of a text element.

As part of data preparation for an NLP model, these elements require further filtering, to mitigate potential detrimental impacts on model efficiency caused by superfluous content. This ensuing phase entails a deliberate omission of specific components, particularly 'Headers' and 'Footers'. As a result, this refinement process retains only four core elements: 'Title', 'Text', 'Image', and 'Table', thereby ensuring a concise and targeted dataset for advanced analysis..

For the "Title" and "Text" elements, prior to integration into NLP models, rigorous data cleaning is essential to avoid efficiency losses caused by extraneous information. To tackle this issue, specialized functions within the 'Unstructured Documentation' cleaning framework are utilized (Unstructured, 2023). These functions effectively merge paragraphs separated by newlines, remove initial bullets and dashes, and eliminate surplus whitespace. This process significantly enhances the textual data's clarity and structural integrity, which is crucial for effective model performance.

For the "Table" elements, the core textual information is retained in the element's 'text attribute'. To preserve the formatting fidelity of these tables, their HTML representation is also stored, specifically within 'element.metadata.text_as_html'. This dual-storage approach is critical for ensuring that the table's structural and visual integrity is maintained in its rendered form.

For the "Image" elements, the 'vision_completion' approach leverages the capabilities of the 'gpt-4-vision-preview' API. This method involves generating specific queries that prompt GPT to provide detailed textual descriptions of images. Once these descriptions are obtained, they are inserted back into the data collection, replacing the positions originally occupied by the images. This process ensures a seamless transition from visual to textual data representation in the dataset..

## 3.3 Chunking

In the 'Unstructured Core Library,' essential for document processing in RAG contexts, the 'chunk_by_title' function is noteworthy for its methodical segmentation of documents into distinct subsections, identifying titles as section markers

(Unstructured, 2023). Notably, it treats elements like tables and images as separate sections. The inclusion of the 'multi-page_sections' parameter is significant, facilitating the formation of multi-page sections that maintain thematic continuity. Unlike common practices, the 'combine_text_under_n_chars' parameter set to zero allows each text piece, regardless of length, to be recognized as an individual section, preserving the document's detailed structure. The default 'new_after_n_chars' parameter relies on the function's internal logic for starting new sections. The 'max_characters' parameter, adjusted to 4096, accommodates larger sections, tailored to the specific requirements of the document structure and content

## 3.4 Vector Database construction

By leveraging OpenAI's "text-embedding-ada-002" model via API, embedding vectors are generated that correspond to specific content. This involves transforming data, initially partitioned into chunks through a preceding chunking process, into vector formats. The utilization of the "text-embedding-ada-002" model is pivotal in enabling large language models to locate content in our dataset that aligns with the given input prompt. The resultant vector data are then stored in Pinecone's vector database, where the feature vectors maintain a dimensionality of 1536. This strategic configuration significantly enhances the database's ability to conduct similarity searches and offers notable advantages in data storage capacity. The application of the "text-embedding-ada-002" model thus integrates OpenAI's advanced natural language processing prowess with Pinecone's efficient vector data management, providing a powerful and versatile solution for text search and analysis purposes.

## 4 Experiments and Discussion

In this segment of the research, we have selected one scholarly papers in Chinese and another in English, along with one book in each language, to evaluate the efficacy of the methodologies employed in this study and the performance of the Retrieval-Augmented Generation (RAG) technique. These papers and books include textual, pictorial, and tabular elements. These two categories represent the predominant forms of publicly released documents at present. Papers are commonly available in an editable PDF format, whereas publicly released books are often found in scanned or image-based PDF formats.The specifics of the documents and books utilized for testing are detailed in Table 1.

## 4.1 Data Processing Results

### 4.1.1 Results of Text Processing Results

The processing results for text information are displayed in Figure 2 and 3, featuring four distinct text blocks from the test papers and books: two in Chinese and two in English. The outcomes are evident in the "Title" and "Cleaned Text" sections. Upon converting all documents to the `.docx` format and applying the prescribed process, the methodology proficiently identifies "Title" across various text types and performs comprehensive text cleaning and organization. This underscores the method's robustness in managing different data structures and multiple languages.

**Table 1 Information of papers and books**

| Type | Title | Page Count | Language |
|------|-------|-----------|----------|
| Paper | Full-scale upgrade activated sludge to continuous-flow aerobic granular sludge Implementing microaerobic-aerobic configuration with internal separators | 12 | English |
| | 提质增效背景下排水管网检测技术的应用与总结 | 8 | Chinese |
| Book | Modelling plastic flows in the European Union value chain | 132 | English |
| | 污水处理设备操作维护问答 | 369 | Chinese |

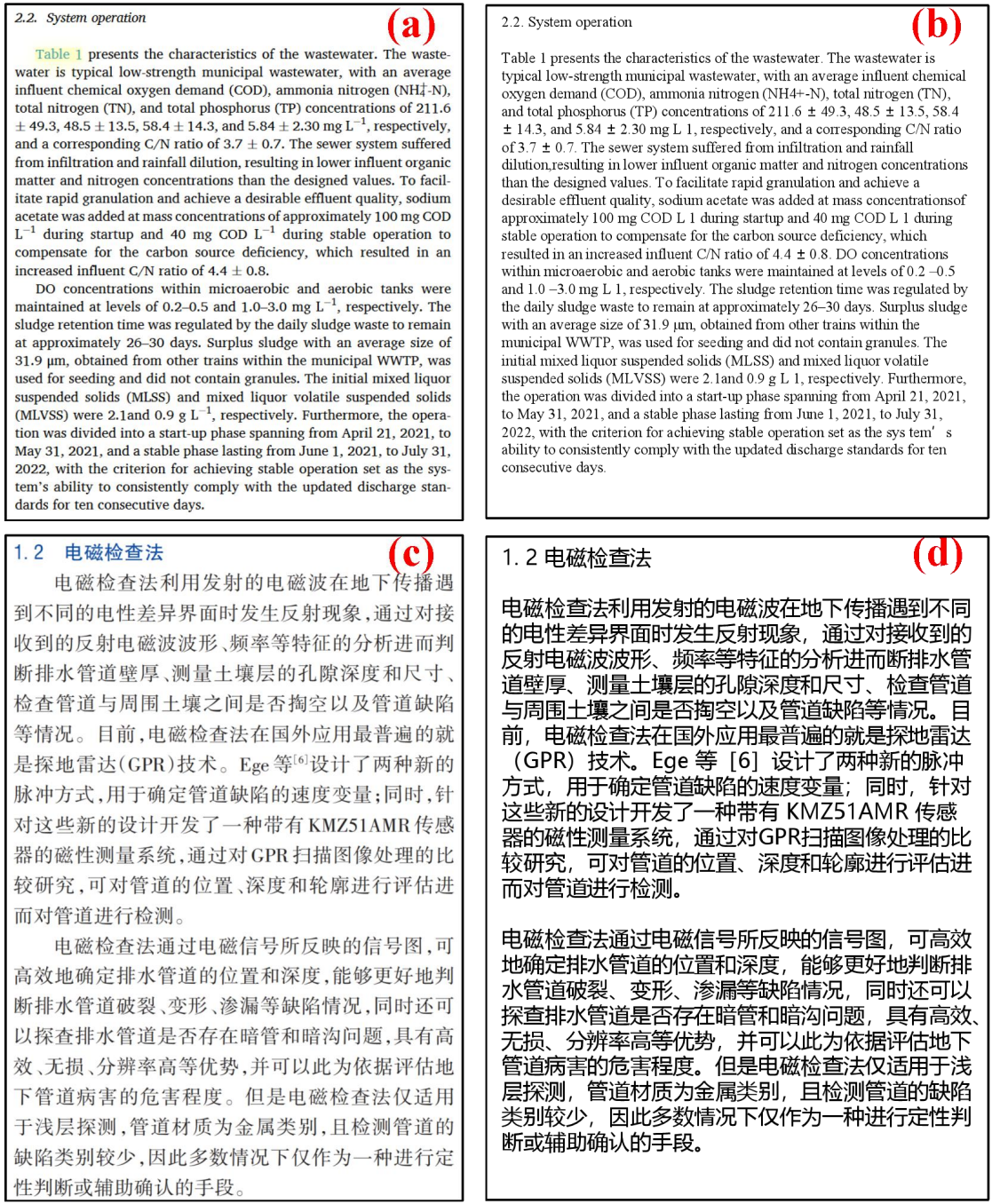

**Fig. 2** Text Processing Results Instances of papers: (a) and (c) are instances of original texts from English and Chinese papers, respectively,while (b) and (d) represent the results of the segmentation into chunks.

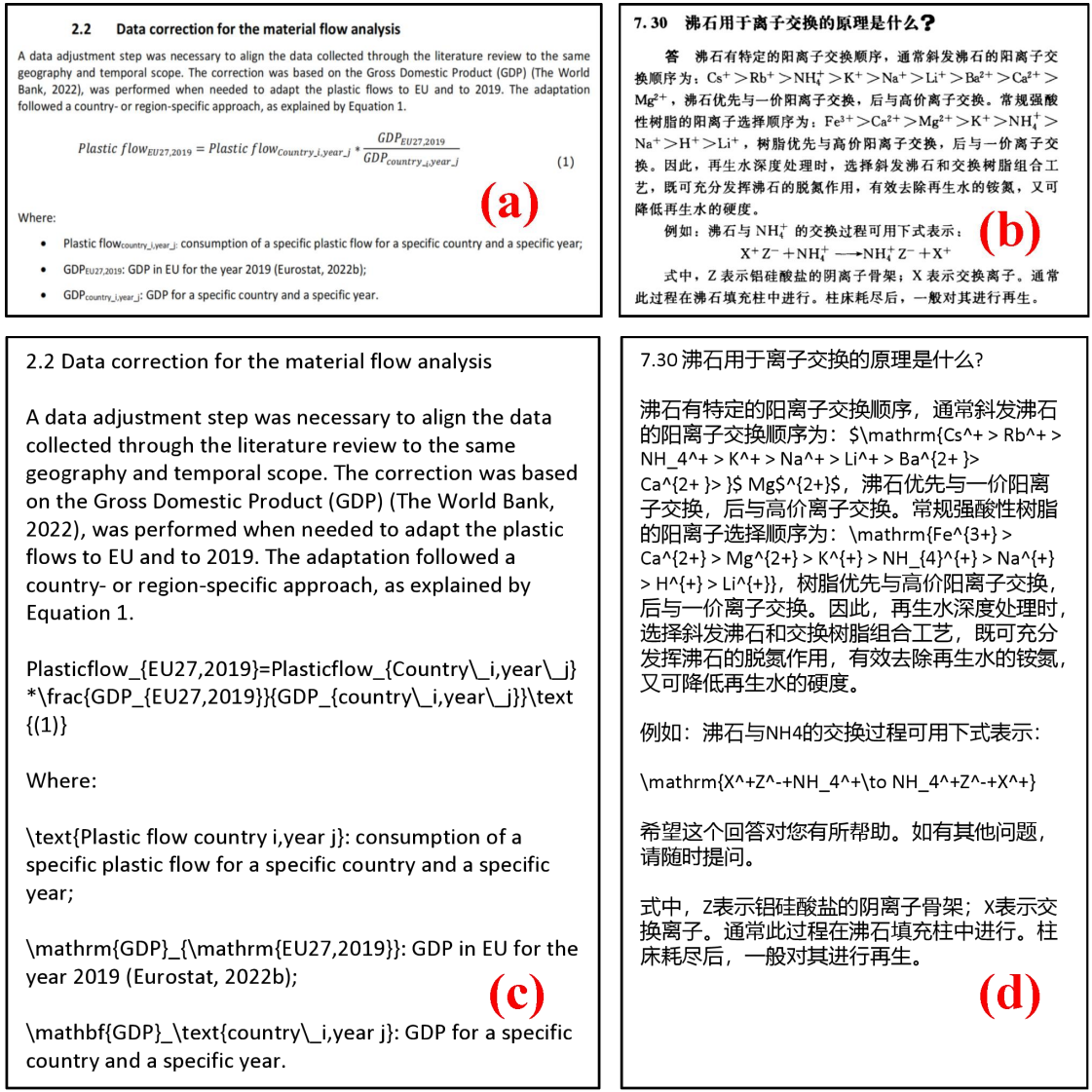

**Fig. 3** Text Processing Results Instances of books: (a) and (b) are instances of original texts from English and Chinese books, respectively,while (c) and (d) represent the results of the segmentation into chunks.

### 4.1.2 Results of Image Processing Results

The results of transforming images into textual descriptions using LLM are presented in Table 2. This research employs an embedding method that leverages the GPT 4.0 LLM to convert images into text, thereby preserving the completeness of the information. The findings indicate that the key information in both English and Chinese images can be effectively extracted. However, due to the model's limited support for Chinese elements, images containing Chinese require additional inputs such as captions or related information to improve the model's recognition accuracy and efficacy, preventing ineffective identifications.

**Table 2 Image processing results**

| NO. | Original Image | Cleaned Text in Chunks |
|---|---|---|
| 1 | 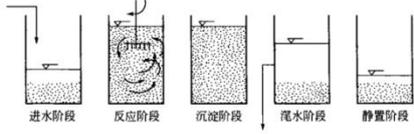 图 3-1 SBR 的基本运行操作过程 | 这张图展示了序批式生物反应器（Sequencing Batch Reactor，简称SBR）的基本运行过程。SBR是一种活性污泥处理过程，经历了几个不同的阶段来处理废水。这些阶段包括：1.进水填充（influent filling）：污水开始填充空的反应器，可以是自流也可以是强制泵送。2.反应（Reaction）：污水在反应器中停留，进行生物化学反应，这个阶段通常包括曝气以促进好氧生物处理。3.沉淀（Settling）：曝气后，搅拌停止让固体物质沉到反应器底部。4.污泥排出（Sludge wasting）：从反应器底部排出沉淀的污泥。5.清水排出（Effluent decanting）：清水从上部被抽出，以完成处理过程，然后开始下一个周期。这个过程周期性重复，能够高效去除污水中的有机物质、氮和磷等污染物 |
| 2 | 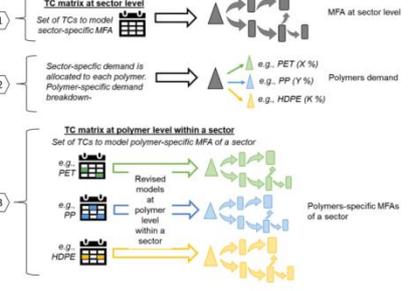 | The diagram illustrates a three-step process for conducting material flow analysis (MFA) at both the sector and polymer levels. Initially, a transformation coefficient (TC) matrix is employed to model the MFA for a specific sector, resulting in an overarching sector-level MFA. Subsequently, the demand within this sector is distributed among various polymers, such as PET, PP, and HDPE, each assigned a percentage of the total demand. Finally, a distinct TC matrix is applied to each polymer type within the sector, refining the models to reflect the individual material flows of PET, PP, and HDPE. These polymer-specific MFAs are depicted through color-coded flow diagrams, providing a detailed visualization of each polymer's flow within the sector, thereby aiding in the optimization of resource use and enhancing recycling efforts. |
| 3 | 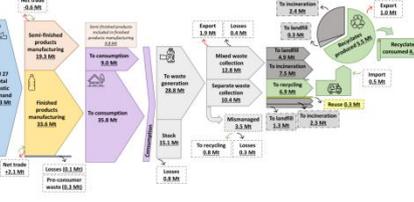 | The image is a detailed flow diagram that represents the lifecycle of plastics within the EU 27, marked by a mix of geometric shapes and a color-coded scheme to depict the journey from production to waste management. Starting with a total plastic demand of 53.3 Mt, it transitions through orange and purple for semi-finished (19.3 Mt) and finished products manufacturing (33.6 Mt). Consumption is shown as 44.8 Mt, leading to a waste generation of 28.8 Mt. The waste management process is highlighted in green, showcasing 1.9 Mt exported and the remainder allocated to incineration, landfill, or recycling, ending with 5.5 Mt of recyclates produced—of which 4.5 Mt is consumed domestically and 1.0 Mt exported, alongside an import of 0.5 Mt of recyclates. Rectangles and arrows indicate flow and quantities, while dashed lines denote net trade figures, exports, imports, and stock. The visual design of the diagram, with annotations for losses and mismanaged waste, underscores the complexities and inefficiencies in the system, providing a clear numeric representation of material flows and inter-process relationships within the plastic economy of the EU. |
| 4 | 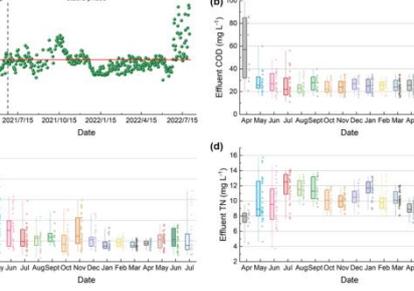 | This composite image presents a detailed visualization of wastewater treatment performance metrics, encapsulating a data set from April 2021 to July 2022. It contains four distinct graphs:(a) A time-series scatter plot depicting the flow rate of wastewater in cubic meters per day, highlighting the operational shift from an initial startup phase to a stable phase, as marked by a vertical dashed line and overlaid with a red trend line to denote mean or target flow rate.(b) A sequential series of box plots illustrating the fluctuating concentrations of effluent Chemical Oxygen Demand (COD) in milligrams per liter, segmented by month for a two-year period, revealing variability and extremes as outliers, represented by individual points outside the upper and lower quartiles.(c) A corresponding set of box plots detailing monthly variations in the concentration of ammonium (NH4+) in the effluent, also in milligrams per liter, showing a generally lower range of values compared to COD, yet similarly depicted with quartile ranges and anomalies as outliers.(d) Lastly, box plots for effluent Total Nitrogen (TN) concentration, presented in the same monthly format and time span, provide insight into the nitrogen levels with a clear graphical representation of the data spread and outliers for each month.This figure collectively portrays the intricacies and dynamics of wastewater treatment parameters, including the stability of flow rates and the monthly variability in key contaminant concentrations, crucial for understanding the efficiency and environmental impact of the treatment process. |

### 4.1.3 Results of Table Processing Results

In the process of data handling, table processing presents significant challenges as tables often contain extensive parameter and comparative analysis information. Such information significantly enhances a LLM's capabilities in data understanding, pattern recognition, and knowledge integration, thereby improving the accuracy and relevance of text generation. In this study, we employed the "text_as_html" method to handle tabular data, with the results displayed in table 3.The corresponding text, rendered as an HTML document, appears as demonstrated in Figure 4.Our analysis indicates that the sections of tables within chunks are expressed in HTML syntax, allowing the saved HTML files to accurately restore the original structure and hierarchy of the tables when opened, ensuring the correct identification and extraction of information.

**Table 3 Table processing results**

| NO. | Original Table | Cleaned text in Chunks |
|---|---|---|
| 1 | Table 1 — Wastewater characteristics (Parameter (mg L⁻¹), Designed influent, Actual influent, Effluent requirements with Startup phase (and before upgrade) and Stable phase; COD 300, 211.6 ± 48.2, 50, 40; NH₄-N, 48.5 ± 13.1, 5 (8), 2 (3.5); TN 70, 58.4 ± 13.9, 15, 15; TP 4, 5.8 ± 2.3, 0.5, 0.4) | Table 1<br>Wastewater characteristics.<br><table><thead><tr><th>Parameter (mg L− 1)</th><th>Designed influent</th><th>Actual influent</th><th>Effluent requirements</th><th>Effluent requirements</th></tr></thead><tbody><tr><td>Parameter (mg L− 1)</td><td>Designed influent</td><td>Actual influent</td><td>Startup phase (and before upgrade)</td><td>Stable phase</td></tr><tr><td>COD</td><td>300</td><td>211.6 ± 48.2</td><td>50</td><td>40</td></tr><tr><td>NH4-N</td><td></td><td>48.5 ± 13.1</td><td>5 (8)</td><td>2 (3.5)</td></tr><tr><td>TN</td><td>70</td><td>58.4 ± 13.9</td><td>15</td><td>15</td></tr><tr><td>TP</td><td>4</td><td>5.8 ± 2.3</td><td>0.5</td><td>0.4</td></tr></tbody></table> |
| 2 | Table 9. Total recyclates production for the sector-specific material flow analysis (expressed as megatonnes) resulting from the sensitivity alternative cases, compared to the results of the 'Base Scenario' sector-specific material flow analysis (calculated as a percentage variation). A description of the rationale for each sensitivity scenario is described in Section 2.3.4. (Note: E = Electrical and Electronic Equipment (EEE)). Columns: Identifier, Sensitivity alternative case name, Recyclates produced (and consumed in EU27) [Mt], Percentage variation with respect to the base MFA [%]. Rows: 1 All manufactured products are consumed 5.88 +32%; 2 Only finished products are sold to end-consumers 3.72 −17%; 3 Reduced stock variation 6.00 +35%; 4 Absence of waste trade 4.71 +6%; 5 Absence of mixed waste collection 9.15 +105%; 6 Absence of mismanaged waste 5.08 +14%; 7 Absence of mismanaged waste being recollected and recycled 3.99 −11%; 8 Revised mismanaged waste assumptions (mismanaged waste only occurs for the transport and EEE sectors and it's not recollected for recycling) 4.04 −10%; 9 Improved recycling performance 6.68 +50% | Table 9. Total recyclates production for the sector-specific material flow analysis (expressed as megatonnes) resulting from the sensitivity alternative cases, compared to the results of the 'Base Scenario' sector-specific material flow analysis (calculated as a percentage variation). A description of the rationale for each sensitivity scenario is described in Section 2.3.4. (Note: E = Electrical and Electronic Equipment (EEE)).<br><table><thead><tr><th>Identifier</th><th>Sensitivity alternative case name</th><th>Recyclates produced (and consumed in EU27)</th><th>Percentage variation with respect to the base MFA [%]</th></tr></thead><tbody>...<tr><td>1</td><td>All manufactured products are consumed</td><td>5.88</td><td>+32%</td></tr>...(continued)...</tbody></table> |
| 3 | 表1 CCTV、QV和声呐的适用性与优缺点 / Tab.1 Applicability, advantages and disadvantages of CCTV, QV and Sonar. Columns: 检测仪器, 适用条件, 优点, 缺点. Rows: CCTV, QV, 声呐 | 表CCTV、QV和声呐的适用性与优缺点<br><table><thead><tr><th>检测仪器</th><th>适用条件</th><th>优点</th><th>缺点</th></tr></thead><tbody><tr><td>CCTV</td>...<td>管道内水位小于管道直径的20%</td>...</tr><tr><td>QV</td>...</tr><tr><td>声呐</td>...<td>管道内水深应大于300 mm</td>...</tr></tbody></table> |

**Table 1** Wastewater characteristics.

| Parameter (mg L− 1) Designed influent Actual influent | | Effluent requirements | Effluent requirements |
|---|---|---|---|
| Parameter (mg L− 1) | Designed influent Actual influent | Startup phase (and before upgrade) | Stable phase |
| COD | 300 | 211.6 ± 48.2 | 50 40 | 50 40 |
| NH-N | 50 | 48.5 ± 13.1 | 5 (8) 2 (3.5) | 5 (8) 2 (3.5) |
| TN | 70 | 58.4 ± 13.9 | 15 15 | 15 15 |
| TP | 4 | 5.8 ± 2.3 | 0.5 0.4 | 0.5 0.4 |

(a)

**Table 9.** Total recyclates production for the sector-specific material flow analysis [expressed as megatonnes] resulting from the sensitivity alternative cases, compared to the results of the 'Base Scenario' sector-specific material flow analysis (calculated as a percentage variation). A description of the rationale for each sensitivity scenario is described in Section2.3.4. (Note: E = Electrical and Electronic Equipment (EEE)).

| Identifier | Sensitivity alternative case name | Recyclates produced (and consumed in EU27) [Mt] | Percentage variation with respect to the base MFA [%] |
|---|---|---|---|
| 1 | All manufactured products are consumed | 5.88 | +32% |
| 2 | Only finished products are sold to end-consumers | 3.72 | -17% |
| 3 | Reduced stock variation | 6 | +35% |
| 4 | Absence of waste trade | 4.71 | +6% |
| 5 | Absence of mixed waste collection | 9.15 | +105% |
| 6 | Absence of mismanaged waste | 5.08 | +14% |
| 7 | Absence of mismanaged waste being recollected and recycled | 3.99 | -11% |
| 8 | Revised mismanaged waste assumptions (mismanaged waste only occurs for the transport and EEE sectors and it's not recollected for recycling) | 4.04 | -10% |
| 9 | Improved recycling performance | 6.68 | +50% |

(b)

表1CCTV、QV和声呐的适用性与优缺点 (c)

| 检测仪器 | 适用条件 | 优点 | 缺点 |
|---|---|---|---|
| CCTV | 管道内水位不大于管道直径的20% | 摄像头可随管道机器人进入管道内部,真实 反映内部情况 | 使用前需进行封堵、清淤、降水工作 |
| QV | 管道内水位不宜大于管的1/2,管段长度不宜大于50 m | 通过摄像头可真实反映管道内部情况;设备 便携,与CCTV相比更加简便 | 检测前需要排除管内积水;只能在 管口进行检测,检测距离短 |
| 声呐 | 管道内水深应大于300 mm | 使用前无需排空积水,适于高水位地区检测 | 仅可以辅助性判断缺陷,管内积水 较少时不适用 |

**Fig. 4 Results of tables elements in chunks converted to html file**

## 4.2 Zero-shot Question Answering Results under RAG

To evaluate the effectiveness of vector knowledge bases constructed using the methodologies outlined in this study for enhancing the expertise of large language models, GPT 4.0 was employed to process the papers and books utilized in this research. A set of fifty questions was randomly generated, focusing on the content of the selected documents. Subsequently, three questions in English and two in Chinese were randomly chosen for testing purposes. GPT 4.0 was then tasked with scoring the responses obtained from these tests, providing an objective measure of the effectiveness of the vector knowledge bases in augmenting the domain-specific knowledge of the language model across different languages. The results of the English and Chinese assessments are presented in Tables 4 and 5, respectively, offering a clear overview of the performance of the vector knowledge bases in enhancing the expertise of GPT 4.0.

**Table 4 Zero-shot question answer results in English**

| NO. | Question and answer | Scores |
|---|---|---|
| 1 | **Question1**：Explain how the "Transfer Coefficients" (TCs) are used to simulate plastic flows in the form of a paragraph? | |

**Answer by GPT 4.0**

Transfer Coefficients (TCs) are crucial parameters in the simulation of plastic flows, particularly in the field of computational fluid dynamics modeling of polymer processing. These coefficients quantify the ease with which molten plastic flows under various conditions, effectively characterizing the relationship between the flow rate of the plastic and the forces applied to it. In simulations, TCs are used to describe how different variables such as temperature, pressure, and material composition affect the flow behavior of plastics. For instance, a higher TC value might indicate that the plastic flows more easily, which can be crucial for processes like injection molding or extrusion. By integrating these coefficients into mathematical models, engineers can predict how plastics will behave in real-world manufacturing scenarios, optimizing processes by adjusting parameters like heat and pressure to achieve desired product characteristics with higher precision and efficiency. This approach helps in reducing material waste, improving product quality, and enhancing the overall efficiency of plastic manufacturing processes.

75/100

**Answer by RAG**

Transfer coefficients (TCs) are essential tools in material flow analysis (MFA) for modeling the flow of materials, including plastics, through different stages of the value chain. These coefficients, expressed as percentages, allocate the flow of a substance from one process to the next, ensuring mass balance is maintained. In the context of plastic flows, TCs are used to estimate the movement of plastics from production to consumption, through waste generation, collection, and management, including recycling and disposal.

For instance, TCs can represent the proportion of plastic waste that is recycled, landfilled, or incinerated. They can also indicate the share of recycled plastics that is used in the production of new products within different sectors. By applying TCs to the total plastic demand or waste generation in a sector, the resulting material flows can be calculated at a granular level, distinguishing between different types of polymers and sectors.

To simulate plastic flows, TCs are first estimated or derived from literature and data sources. These coefficients are then applied to the total plastic mass at each stage of the value chain. For instance, the total plastic waste generated is multiplied by the TC for recycling to find out how much of that waste is recycled. Similarly, the amount of recycled plastic going into new products in a specific sector is calculated by applying the relevant TC.

Overall, TCs enable the detailed modeling of plastic flows, providing insights into how plastics move through different stages and sectors. This information is crucial for understanding and improving the sustainability of plastic management and recycling practices.

95/100

| NO. | Question and answer | Scores |
|-----|---------------------|--------|

**Question2**： Which predefined scenarios showed the greatest potential improvement when assessing the 2025 plastic recycling targets?

### Answer by GPT 4.0

As of my last update, specific predefined scenarios showing the greatest potential improvement for the 2025 plastic recycling targets are not detailed in my training data. However, I can outline typical scenarios that are often considered in strategic planning and policy formulation aimed at improving recycling rates and achieving sustainability goals. These scenarios generally focus on innovations in technology, policy adjustments, and behavioral changes.

High-Potential Scenarios for Plastic Recycling Improvements
Policy-Driven Scenarios:

Mandatory Recycling and Minimum Recycled Content Laws: Policies that require the use of recycled plastic in certain products or mandate recycling can drive demand for recycled materials and ensure steady supply chains.
Extended Producer Responsibility (EPR): Policies that hold producers responsible for the lifecycle of their products, including disposal, can incentivize them to design products that are easier to recycle and to contribute financially to recycling infrastructure.
Technological Advancement Scenarios:

Enhanced Mechanical Recycling: Development and deployment of more efficient mechanical recycling technologies that can handle a broader range of plastics with better output quality.
Scaling Chemical Recycling: Chemical recycling technologies, which break down plastics to their chemical components, could be scaled up to handle mixed or contaminated plastics that are currently non-recyclable.
Market Development Scenarios:

Development of New Markets for Recycled Plastics: Encouraging the use of recycled plastics in higher-value applications, which can be facilitated by improving the quality of recycled resins and creating certifications for recycled content.
Incentives for Using Recycled Content: Financial incentives such as tax breaks or subsidies for products made with recycled plastic can stimulate market demand.
Behavioral and Systemic Change Scenarios:

Consumer Awareness Campaigns: Enhancing consumer awareness about the benefits of recycling and the proper ways to recycle can improve the quality and quantity of collected recyclable materials.
Global Alignment on Plastic Handling: International agreements on managing plastic waste could standardize recycling practices and reduce the export of waste to countries with less robust environmental protections.
Impact on 2025 Targets
These scenarios are designed to tackle different aspects of the plastic lifecycle, from production and use to disposal and recycling. By implementing a combination of these strategies, significant progress can be made towards meeting the 2025 plastic recycling targets, which generally aim for higher recycling rates, reduction in virgin plastic use, and a decrease in plastic waste and pollution. Each scenario not only supports these direct targets but also contributes to broader environmental, economic, and social goals, such as reducing carbon emissions, creating green jobs, and promoting sustainable consumer practices.

For specific details on which scenarios are being prioritized or have shown the most promise based on empirical data or projections as of 2025, it would be beneficial to look at recent reports from relevant environmental agencies, government bodies, or international organizations involved in waste management and recycling initiatives.

60/100

### Answer by RAG

The predefined scenarios that showed the greatest potential improvement when assessing the 2025 plastic recycling targets are as follows:

1. **Improved Waste Collection Scenario (Scenario B):** This scenario, which involves a 30% increase in separate collection for packaging and a 10% increase for other sectors, leads to the highest estimated recyclates consumption of 6.47Mt by 2025, representing a 45% increase compared to 2019.

2. **Combined Scenario (Scenarios F1, F2, F3):** When all actions are combined, including a reduction in waste export, improved waste collection, improved sorting yields, and improved recycling yields, along with variations in plastic production (+10%, -10%, 0%), the EU27 target could be reached and surpassed by 2025. Specifically, the combined scenario with a +10% variation in production could achieve 11.13Mt of recyclates consumed by 2025, with an overall recyclates consumption rate of 35.5%.

3. **Improved Management of Separately Collected Plastic Waste (Scenario C):** Assuming 15% more plastic packaging waste separately collected is sent to recycling facilities, and 10% more for other sectors, this scenario could result in 5.23Mt of recyclates being consumed in the EU27 and an overall recyclates consumption rate of 19.5%.

4. **Improved Recycling Performance (Scenario D):** Assuming a 20% lower rejects from recycling facilities for packaging waste and 10% for other sectors, this scenario could achieve 5.57Mt of recyclates being consumed in the EU and a recyclates consumption rate around 20.8%.

In summary, the combined scenario shows the highest potential for achieving the EU/CPA target of 8.8Mt of plastics recyclates by 2025, with the improved waste collection scenario also demonstrating significant potential.

95/100

| NO. | Question and answer | Scores |
|---|---|---|

**Question3：** How did the microaerobic-aerobic configuration impact the microbial community structure and pollutant removal pathways?

### Answer by GPT 4.0

The microaerobic-aerobic configuration in wastewater treatment systems, particularly those using Aerobic Granular Sludge (AGS) technology, has a significant impact on both the microbial community structure and the pollutant removal pathways. This configuration typically involves a sequential arrangement where wastewater first passes through a microaerobic zone before entering an aerobic zone. This setup is designed to optimize both biological nutrient removal and the operational stability of the system. Here's how this configuration influences microbial dynamics and treatment processes:

Impact on Microbial Community Structure
Enhanced Microbial Diversity: The microaerobic zone provides a unique environment with low oxygen levels that is conducive to the growth of facultative anaerobic microorganisms, which can survive in both aerobic and anaerobic conditions. This promotes a more diverse microbial community compared to a purely aerobic system.

Selective Pressure for Denitrifiers: The low-oxygen conditions in the microaerobic zone favor denitrifying bacteria that can reduce nitrate to nitrogen gas, thus enhancing denitrification. This is crucial for the removal of nitrogen from wastewater.

Growth of Phosphorus-Accumulating Organisms (PAOs): PAOs can thrive under alternating anaerobic and aerobic conditions, which are effectively provided by the microaerobic followed by aerobic zones. These organisms uptake phosphorus under aerobic conditions and release it under anaerobic conditions, aiding in biological phosphorus removal.

Reduction of Bulking and Foaming: The configuration can help in controlling filamentous bacteria, which are often responsible for sludge bulking and foaming in purely aerobic systems. By adjusting the oxygen levels, the growth of filamentous bacteria can be managed, improving sludge settling characteristics.

Impact on Pollutant Removal Pathways
Enhanced Nitrogen Removal: The microaerobic zone allows for partial denitrification, which can be completed in the aerobic zone where nitrification occurs primarily. The sequential arrangement ensures more effective nitrogen removal as compared to systems with only aerobic processes.

Improved Phosphorus Removal: The alternation between microaerobic and aerobic conditions is ideal for cyclic phosphorus uptake and release by PAOs. This enhances the biological removal of phosphorus, reducing the need for chemical precipitation.

Organic Matter Degradation: The presence of a microaerobic zone can aid in the breakdown of certain types of organic compounds that are more readily processed under low oxygen conditions, potentially leading to a more efficient overall reduction of biochemical oxygen demand (BOD) and chemical oxygen demand (COD).

Pathogen Reduction: Some pathogens are more effectively controlled under fluctuating aerobic and anaerobic conditions, which could potentially reduce the pathogen load in the effluent.

Operational Stability and Flexibility: This configuration allows for more flexible operation and better adaptation to varying loads and influent characteristics, which can improve the overall stability and efficiency of the wastewater treatment process.

The microaerobic-aerobic configuration essentially creates a more controlled and tailored environment for wastewater treatment, optimizing microbial community development for specific pollutant removal tasks and improving the resilience and efficiency of the system. This configuration is particularly advantageous in AGS technology where the physical characteristics of the granules also help to maintain distinct zones with different oxygen concentrations, further enhancing process performance.

**85/100**

### Answer by RAG

The microaerobic-aerobic configuration significantly impacted the microbial community structure and pollutant removal pathways in the following ways:

1. **Microbial Community Shifts:** The transition from flocculent-activated sludge to well-defined aerobic granular sludge (AGS) was observed, indicating a shift in the microbial community. The dominant phyla remained Proteobacteria, Bacteroidota, and Chloroflexi, but their relative abundances varied. Notably, the relative abundances of denitrifying bacteria (DNB) such as Saprospiraceae, Caldilineaceae, and Comamonadaceae increased, indicating a shift towards DNB enrichment.

2. **Pollutant Removal Pathways:** The primary removal of pollutants occurred within the microaerobic tanks. Simultaneous nitrification, denitrification, and phosphorus removal were observed, with the microaerobic tanks accounting for 77-93% of the removal of COD, NH-N, TN, and TP, respectively. The enrichment of denitrifying phosphorus-accumulating organisms (DPAOs) like Hydrogenophaga, Accumulibacter, Azospira, Dechloromonas, and Pseudomonas provided an essential microbial foundation for these processes.

3. **Granulation Mechanisms:** The incorporation of internal separators in the aerobic tanks induced shifts in the flow pattern, transitioning from a single-circulation cell to multiple voritcal cells. This alteration amplified the local velocity gradients, generating the required shear forces to drive granulation. Additionally, the microaerobic and aerobic tanks operated under feast and famine conditions, creating a microbial selection pressure that favored granulation.

In summary, the microaerobic-aerobic configuration induced shifts in the microbial community towards DNB enrichment, facilitated simultaneous nitrification-denitrification and phosphorus removal primarily in the microaerobic tanks, and induced granulation through altered flow patterns and microbial selection pressures. This configuration effectively upgraded the existing activated sludge system into a continuous-flow AGS system with enhanced pollutant removal capabilities.

**95/100**

**Table 5 Zero-shot question answer results in Chinese**

| NO. | Question and answer | Scores |
|---|---|---|

**Question1**：几种常用树脂再生剂的适用对象、浓度范围及相对用量是多少？

**Answer by GPT 4.0**

树脂再生剂用于再生离子交换树脂，以恢复其交换容量。常用的树脂再生剂主要包括酸和碱性化学品，用于阳离子树脂和阴离子树脂的再生。下面详细介绍这些再生剂的适用对象、浓度范围及相对用量：

1. **硫酸钠（NaHSO₄）**
   - **适用对象**：用于阳离子树脂的再生。
   - **浓度范围**：一般使用2-4%的溶液。
   - **相对用量**：大约为树脂体积的2-3倍。
2. **盐酸（HCl）**
   - **适用对象**：主要用于阳离子树脂的再生，特别是在处理含硬度较高的水质时。
   - **浓度范围**：通常使用4-10%的溶液。
   - **相对用量**：约100-150克HCl（以100%计）/L树脂。
3. **氢氧化钠（NaOH）**
   - **适用对象**：用于阴离子树脂的再生。
   - **浓度范围**：通常使用2-4%的溶液。
   - **相对用量**：约80-120克NaOH（以100%计）/L树脂。
4. **氯化钠（NaCl）**
   - **适用对象**：阳离子树脂的再生，适用于软化水处理中。
   - **浓度范围**：通常使用8-10%的盐水溶液。
   - **相对用量**：约为树脂体积的2-3倍。
5. **硝酸钠（NaNO₃）**
   - **适用对象**：适用于特定的阳离子树脂再生，尤其是在某些工业应用中。
   - **浓度范围**：通常使用1-5%的溶液。
   - **相对用量**：相对较少，依应用而定。

80/100

**Answer by RAG**

几种常用树脂再生剂的适用对象、浓度范围及相对用量如下：

1. HCl：
   - 适用对象：强酸性，H型阳离子交换树脂
   - 浓度范围：3-9%
   - 相对用量：3-5倍理论用量
   - 适用对象：弱酸性，H型阳离子交换树脂
   - 浓度范围：4-10%
   - 相对用量：1.5-2倍理论用量
   - 适用对象：强碱性，Cl型阴离子交换树脂
   - 浓度范围：8-12%
   - 相对用量：4-5倍理论用量
   - 适用对象：弱碱性，Cl型阴离子交换树脂
   - 浓度范围：8-12%
   - 相对用量：1.5-2倍理论用量
2. NaCl：
   - 适用对象：强酸性，Na型阳离子交换树脂
   - 浓度范围：8-10%
   - 相对用量：3-5倍理论用量
3. NaOH：
   - 适用对象：弱酸性，Na型阳离子交换树脂
   - 浓度范围：4-6%
   - 相对用量：1.5-2倍理论用量
   - 适用对象：强碱性，OH型阴离子交换树脂
   - 浓度范围：4-6%
   - 相对用量：4-5倍理论用量
   - 适用对象：弱碱性，OH型阴离子交换树脂
   - 浓度范围：3-5%
   - 相对用量：1.5-2倍理论用量
4. NH₄OH：
   - 适用对象：弱碱性，Cl型阴离子交换树脂
   - 浓度范围：3-5%
   - 相对用量：1.5-2倍理论用量

95/100

| NO. | Question and answer | Scores |
|---|---|---|

**Question2**：在排水管网检查中电磁检查法有哪些应用案例？

### Answer by GPT 4.0

电磁检查法在排水管网检查中是一个有用的非破坏性检测技术，尤其适用于识别金属管道的腐蚀、磨损和其他结构缺陷。这种方法通常涉及使用特定设备，如电磁探测器和相关的传感器系统，来评估管道的完整性。以下是一些电磁检查法在排水管网检查中的应用案例：

1. **检测金属管道的腐蚀和磨损**：
   - 在金属管道中，电磁检查技术可以用来探测腐蚀产生的金属厚度减少。这种检查可以提供有关管道内外壁腐蚀程度的详细信息，帮助决定是否需要修复或更换管道。
2. **确定管道壁的厚度**：
   - 使用电磁波测量金属管道的厚度，尤其是在管道外部难以直接测量的情况下。这有助于监测老化管网的状况，预测可能的破裂风险。
3. **定位地下金属管道**：
   - 在城市或复杂环境中，电磁探测器可以用来精确地定位地下金属排水管道的位置。这对于维护工作和避免在施工中意外损坏管道至关重要。
4. **评估管道接头和焊缝的完整性**：
   - 电磁检查也可以用于评估管道接头和焊缝区域的结构完整性。这是检查老化管网中潜在弱点的重要环节。
5. **铁磁性管道的异常检测**：
   - 对于铁磁性材料的管道，电磁检查技术可以识别因内部应力或外部影响而导致的磁场异常，这些异常可指示存在裂纹或其他结构问题。

在实际应用中，例如，在某些城市的大规模排水管网改造项目中，电磁检查法被用来评估整个管网的状况，以优化修复和更换计划。这种技术为管网的长期运行提供了一种有效的监测和预防性维护手段。

虽然这些案例提供了电磁检查法的一般应用示例，具体实施细节和结果可能会根据项目、地点和使用的具体技术而有所不同。

**75/100**

### Answer by RAG

电磁检查法在排水管网检查中的应用案例主要包括以下几个方面：

1. **探地雷达（GPR）技术**：通过发射电磁波并分析其反射波，可以检测排水管道的位置、深度、轮廓以及管道缺陷等情况。例如，Ege等设计了新的脉冲方式，用于确定管道缺陷的速度变量，并开发了一种带有KMZ51AMR传感器的磁性管道检测系统，通过GPR扫描图像处理的比较研究，对管道进行评估和检测。
2. **电磁信号分析**：电磁检查法通过分析电磁信号所反映的信号图，可以高效地确定排水管道的位置和深度，判断管道破裂、变形、渗漏等缺陷情况，并探查是否存在暗管和暗沟问题。
3. **局限性**：电磁检查法适用于浅层探测，管道材质为金属类别，且检测管道的缺陷类别较少，因此多数情况下仅作为一种进行定性判断或辅助确认的手段。
4. **结合其他方法**：在实际应用中，电磁检查法常与影像检查法、水质水量检查法相结合，以获得更全面的检测效果。例如，陈泽鑫等在珠海横琴新区污水系统诊断工作中，同步采用水质水量检测法定性定量分析管网现状问题，结合电磁检查法定位管道缺陷病害问题，运用系统思维分析出厂污水浓度偏低的主要原因。

综上所述，电磁检查法在排水管网检查中具有高效、无损、分辨率高等优势，但也存在一定的局限性，需要与其他检测方法相结合，以获得更准确的检测结果。

**90/100**

The results presented in this study provide compelling evidence that vector knowledge bases constructed using the methodologies described herein can significantly enhance the ability of large language models to acquire and apply domain-specific information. This improvement is manifested across several critical dimensions, including clarity, specificity, accuracy, technical depth, and comprehensiveness. By effectively augmenting the knowledge acquisition process, these vector knowledge bases enable language models to generate responses of substantially higher quality, demonstrating their efficacy in improving the performance of large language models in specialized domains. These findings underscore the potential of vector knowledge bases as a powerful tool for enhancing the accuracy and relevance of language model outputs in domain-specific contexts, paving the way for more effective and efficient natural language processing applications in various specialized fields.

## Conclusion

The methodologies developed in this study significantly enhance the capability of LLMs to leverage domain-specific knowledge through the construction of vector knowledge bases. Our experiments demonstrate the effectiveness of the RAG approach, where LLMs, equipped with these bases, show substantial improvements in generating precise, relevant, and contextually rich responses. This advancement is particularly evident in the environmental science and wastewater treatment sectors, where the integration of vector databases enables the detailed understanding and management of complex data. The successful application of these methods promises a broader utility of LLMs, paving the way for more sophisticated natural language processing applications in various specialized fields. This research not only validates the feasibility of enhancing LLMs performance with structured vector databases but also sets a foundation for future innovations in AI-driven data processing and analysis in environmental engineering.